\documentclass[aps,pre,twocolumn,superscriptaddress,showkeys,amsmath,amssymb,footinbib,longbibliography]{revtex4-2}
\usepackage[english]{babel}
\usepackage{amssymb,amsbsy,amsmath}
\usepackage{graphicx}
\usepackage{dcolumn}
\usepackage{bm}
\usepackage{subfigure}
\usepackage{color}
\usepackage{textcomp}
\usepackage{comment}
\usepackage{mathrsfs}
\usepackage[colorlinks,bookmarks=false,citecolor=blue,linkcolor=red,urlcolor=blue]{hyperref}

\newcommand{\xref}[1]{\protect\ref{#1}}
\newcommand{\figref}[1]{Fig.~\protect\ref{#1}}

\def\bra#1{\langle \, {#1} \, | \,}
\def\ket#1{\, | \, {#1} \, \rangle}

\renewcommand{\eqref}[1]{Eq.~(\protect\ref{#1})}

\def\rket#1{\, | \, {#1} \, )}
\def\rbra#1{( \, {#1} \, | \,}
\newcommand{\rbraket}[2]{( \, {#1} \, | \, {#2} \, )}

\newcommand{\changed}[1]{#1}

\begin{document}
%
\title{Escaping the Krylov space during finite-precision Lanczos}

\author{Jannis Eckseler}
\email{jeckseler@physik.uni-bielefeld.de}
\author{Max Pieper}
\email{mpieper@physik.uni-bielefeld.de}
\author{J\"urgen Schnack}
\email{jschnack@uni-bielefeld.de}
\affiliation{Fakult\"at f\"ur Physik, Universit\"at Bielefeld, Postfach 100131, D-33501 Bielefeld, Germany}

\date{\today}

\begin{abstract}
The Lanczos algorithm, introduced by Cornelius Lanczos, has been known for a long time and is widely used in computational physics. 
While often employed to approximate extreme eigenvalues and eigenvectors of an operator, 
recently interest in the sequence of basis vectors produced by the algorithm rose
in the context of Krylov complexity. Although it is generally accepted and partially proven
that the procedure is numerically stable for approximating the eigenvalues, 
there are numerical problems when investigating the Krylov basis constructed via
the Lanczos procedure. In this paper, we show that loss of orthogonality 
and the attempt of reorthogonalization fall short of understanding and addressing the problem. 
Instead, the sequence of numerical Lanczos vectors in finite-precision arithmetic 
escapes the true vector space spanned by the exact Lanczos vectors.
This poses the real threat to an interpretation in view of
the operator growth hypothesis.
\end{abstract}

\keywords{Krylov complexity, Lanczos algorithm}

\maketitle

\section{Introduction}
\label{sec-1}

The Lanczos procedure \cite{Lan:JRNBS50} is applied in many contexts in physics 
for instance to approximate extreme eigenvalues of operators such as ground states 
and low-lying excited states of Hamiltonians
as well as functions of the Hamiltonian such as the partition function, compare e.g.\ \cite{Ski:88,Hut:CSSC89,DrS:PRL93,JaP:PRB94,ADE:PRB03,WWA:RMP06,AvT:ACM11,SAI:NM17,Chen:A24}.
Its accuracy has been thoroughly investigated over time with the general result that the method 
is very accurate and robust in finite precision arithmetics 
which is an important issue for computer applications \cite{SIMON1984101,book:8990,Wue:SJMAA05,Wue:BNM05,MeS:AN06,RoA:FCM15,SRS:PRR20,CGM:SIAM22,ChM:ETNA24}.

Recent efforts to understand the emergence of thermodynamics in closed quantum systems under
unitary time evolution \cite{Deu:PRA91,Sre:PRE94,ScF:NPA96,ScF:PLB97,Tas:PRL98,RDO:N08,Rei:PRL08,PSS:RMP11,ReK:NJP:12,ShF-NJP:12,SKN:PRL14,GoE:RPP16,AKP:AP16,BIS:PR16,STS:QST18,ReG:PA19,PhysRevX.9.041017}
produced a variety of measures to discriminate between, e.g.,
integrable and quantum chaotic behavior such as \emph{level statistics} (Poisson vs.\ Wigner-Dyson)
and \emph{Thouless time} to mention just two of them \cite{STS:PRB19,LVG:PRL20}.
A new approach to separate quantum chaotic dynamics from the dynamics of integrable
quantum systems is given by the \emph{universal operator growth hypothesis} \cite{PhysRevX.9.041017} 
which heavily relies on the Lanczos procedure and thus sparked new and intensified interest 
in the Lanczos algorithm applied to finite size systems as well as to systems in the thermodynamic 
limit \cite{PhysRevE.106.014152,PhysRevResearch.4.013041,HMT:JHEP23,RSS:JHEP22,PhysRevB.111.165106,NMB:JPCM24,
BNN:JHEP23,NMM:PR25,San:A24,PhysRevD.109.066010,MeJ:PRB24,BBC:A25}.
The hypothesis links the asymptotic behavior of the Lanczos coefficients to the characteristic 
of integrability of the system. However in the process of calculating these coefficients, 
numerical difficulties can occur. 

In our investigation, we show that the numerical sequence of eigenvectors 
in finite precision arithmetic escapes the true vector space spanned 
by the exact Lanczos vectors. The underlying reason is that this space is embedded 
in the much larger computational vector space. We think that loss of orthogonality 
to previous Lanczos vectors is a minor problem compared to acquiring components into
the Lanczos sequence that do not belong to the true space. These components grow 
exponentially fast which is a genuine property of the Lanczos procedure. 
This numerical problem cannot be cured by reorthogonalization.
Therefore, the interpretation of numerical Lanczos sequences 
in view of the operator growth hypothesis
appears questionable to us.

The paper is organized as follows: In Section \xref{sec-2} we recapitulate essentials
of the Lanczos procedure, and in Section \xref{sec-3} we discuss and demonstrate 
with a few examples of finite quantum spin systems 
how the mentioned numerical problems arise. The paper closes with a summary. 
In the appendix, we demonstrate with two examples that the identified problems 
also arise in ordinary Lanczos procedures for ket states or wave functions.

\section{Lanczos algorithm and the operator growth hypothesis}
\label{sec-2}

Since the operator growth hypothesis makes a statement about the Lanczos coefficients 
we first have to review the Lanczos algorithm \cite{Lan:JRNBS50}. 
To start the Lanczos algorithm one needs a linear operator $H$ 
and a normalized starting vector $\ket{\psi}$ or, 
when considering the Hilbert space of operators,
a starting operator $\rket{O}$ and a superoperator $\mathcal{L}$ \cite{Muus1972}. 
This Hilbert space of operators has to be provided with an inner product, 
which in this paper is chosen to be the infinite temperature product 
$\rbraket{A}{B}=Tr(A^{\dagger}B)/dim(\mathcal{H})$, also known as the Frobenius product, 
with the induced norm $||A||=\sqrt{\rbraket{A}{A}}$. Note that, 
since the Hilbert space of operators is also a vector space, 
we might call operators vectors as well. 
However, the distinction between vectors in the ket-vector space 
$\mathcal{H}$ and the {operator-vector} space $\mathcal{H} \otimes \mathcal{H}$ 
will be made clear by different notations as $\ket{v}$ and $\rket{w}$, respectively. 
For physical systems the linear superoperator of interest is the Liouvillian $\mathcal{L}$
which is defined via $\mathcal{L}\rket{O}=[H,O]$ where $H$ is the Hamiltonian of the system.
To start the Lanczos algorithm we set $\rket{O_0}=\rket{O}$, 
$b_1=||\mathcal{L}O_0||$, and $\rket{O_1}=b_1^{-1}L\rket{O_0}$. 
We then follow the iteration scheme
\begin{align}
\label{operatorlanczos}
     \rket{A_n}= & \mathcal{L} \rket{O_{n-1}} - b_{n-1} \rket{O_{n-2}} \\
      b_n= & ||A_n|| \\
     \rket{O_n}= & b_n^{-1} \rket{A_n}
\end{align}
to construct the Krylov basis $\{ \rket{O_n} \}$, which is an ONB of the Krylov space.
Note that since we are considering a starting vector $\rket{O}$, 
which is a hermitian operator, one only needs to project out the second latest $\rket{O_n-2}$, 
because the inner product $\rbraket{O_n-1}{\mathcal{L}O_n-1}$ vanishes for hermitian and skewhermitian operators $O$.

In exact arithmetic the Lanczos algorithm works just as described, however with finite numerical precision the constructed $\rket{O_n}$ are generally not an ONB. While for many applications of the Lanczos algorithm this does not pose a problem, it is crucial in our case, since it can heavily influence the Lanczos sequence. A usual attempt at trying to better the results is to reorthogonalize with respect to all prior operators, so all $\rket{O_k}$ with $k<n$ or even repeat the reorthogonalization procedure a second time as suggested in \cite{RSS:JHEP21}, which we call double reorthogonalization.
While this may lead to termination of the Lanczos sequence in some cases, there are still numerical problems, 
which will be discussed in section \ref{sec-3}.

The Liouvillian in the Krylov basis is a tridiagonal matrix of the form
\begin{align}
    \mathcal{L} = \begin{pmatrix}{}
    0 & b_1 & 0   & 0   & \cdots & \cdots & 0\\
    b_1 & 0 & b_2 & 0   & \cdots & \cdots & 0\\
    0   & b_2 & 0 & b_3 & \cdots & \cdots & 0\\
    \vdots & &  \ddots & \ddots & \ddots & &  \vdots\\
    0 & \cdots & \cdots & b_{m-3} & 0 & b_{m-2} & 0 \\
    0 & \cdots & \cdots & 0       & b_{m-2} & 0 & b_{m-1} \\
    0 & \cdots & \cdots & 0       & 0       & b_{m-1}     & 0 
    \end{pmatrix}\label{eq:lanc vector matrix} \quad ,
\end{align}{}
with $\mathcal{L}_{mn}=\rbra{O_m} \mathcal{L} \rket{O_n}$.
The Lanczos coefficients $b_n$ are the quantity of interest for the universal operator growth hypothesis, 
which states that for infinite nonintegrable systems 
and a local starting operator $O$
these $b_n$ should grow asymptotically linearly with $n$.

\section{Numerical problems in finite systems}
\label{sec-3}

While originally formulated for infinite systems, there has also been interest in the 
behavior of the Lanczos sequence in finite systems. 
Obviously, the universal operator growth hypothesis cannot hold for finite systems 
since the Lanczos sequence will terminate at some point and can therefore not grow asymptotically linearly.
However one can still examine the Lanczos sequence up to that point and compare the behavior in different systems \cite{Barbon2019,RSS:JHEP21}.

It is known that the dimension of the Krylov space 
$\mathcal{K}(\mathcal{L},O)=span \{ \rket{O}, \mathcal{L} \rket{O}, \mathcal{L}^2 \rket{O},... \}$ 
is the minimum number of eigenvectors 
of $\mathcal{L}$
needed to represent the starting vector
$\rket{O}$. 
Therefore, if the system is small enough to diagonalize it computationally, 
one can predict the actual Krylov space dimension. 
This is a useful insight to show that numerically the Lanczos sequence
does not terminate at the edge of the Krylov space. 
In exact arithmetic the $b_n$ should vanish for 
$n_{\text{max}}=dim(\mathcal{K}(\mathcal{L},O))+1$ 
since at $n_{\text{max}}-1$ 
we already constructed a full ONB for the Krylov space at hand. 
Iterating one more step should yield an operator $A_n$ with $||A_n||=0$, hence $b_n=0$.

\subsection{Tilted-field Ising ring \label{sec:isring}}

To numerically investigate this behavior we considered the tilted-field Ising model 
given by the Hamiltonian
\begin{align}
    H=\sum _{i=0}^{N-1} s_i^z s_{i+1}^z +h_z s_i^z + h_x s_i^x
    \ ,
\end{align}
where we used periodic boundary conditions and chose $N=6$, $h_z=1$ as well as $h_x=-1.05$.
The starting vector for which the Lanczos method is applied was chosen to be the 1-local operator 
${s}_0^z$.

In order to compute the dimension of the Krylov space $\mathcal{K}(\mathcal{L}, {s}_0^z)$ 
one first has to find the eigenvectors $\rket{E_k^{\mathcal{L}}}$ 
and eigenvalues $E_k^{\mathcal{L}}$ of $\mathcal{L}$. 
This can either be done by diagonalizing $\mathcal{L}$ itself or by diagonalizing $H$ 
to obtain its eigenvalues $E_m^H$ and eigenvectors $\ket{E_m^H}$,
and using these to construct an eigenvector and eigenvalue of $\mathcal{L}$ via
\begin{align}
\label{eq:ham-liou}
    \rket{E_{k(m,n)}^{\mathcal{L}}}=\ket{E_m^H}\bra{E_n^H} \ \text{and} \   E^{\mathcal{L}}_{k(m,n)}=E^H_m-E^H_n,
\end{align}
where $k(m,n)$ is the integer enumeration for the eigenstates of the Liouvillian

Now let $\{ E^{\mathcal{L}}_p \}$ be the set of eigenspaces of $\mathcal{L}$.
Then the starting vector $\rket{s_0^z}$ can be decomposed as
\begin{align}
    \rket{s_0^z}=\sum_{p=1}^{\#\{ E^{\mathcal{L}}_p \}} d_p 
    \sum_{k=1}^{dim(E^{\mathcal{L}}_p)} c_k^p \rket{E^\mathcal{L}_{k,p}}
    \ .
\end{align}
One can represent the second sum as one combined vector $ \rket{\bar{E}_p^\mathcal{L}}=\sum_{k=1}^{dim(E^{\mathcal{L}}_p)} c_k^p \rket{E^\mathcal{L}_{k,p}}$, which are the only eigenvectors needed to represent our starting vector in this eigenspace, 
such that we can now write
\begin{align}
    \rket{s_0^z}=\sum_{p=1}^{\#\{ E^{\mathcal{L}}_p \}} d_p \rket{\bar{E}_p^\mathcal{L}}
    \ .
\end{align}
As mentioned before, the dimension of $\mathcal{K}(\mathcal{L},s_o^z)$ is just the 
number of eigenvectors to different eigenvalues that $\rket{s_0^z}$ has an overlap with, 
so the number of coefficients $d_p$ with $d_p \neq 0$. 

\begin{figure}[ht!]
\centering
\includegraphics*[width=0.9\columnwidth]{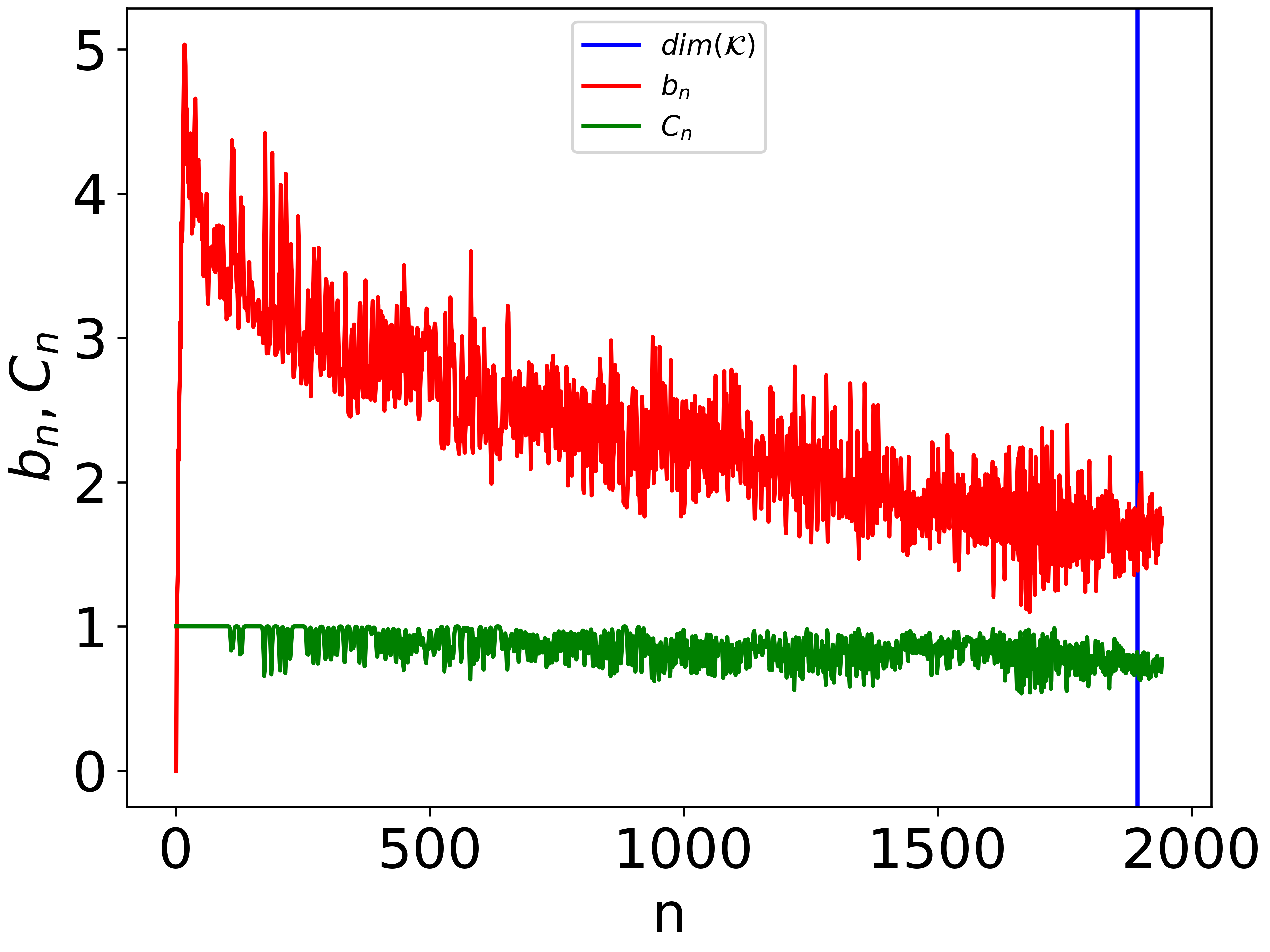}
\caption{Lanczos sequence $b_n$ with double reorthogonalization and overlap $C_n$ for the tilted-field Ising 
ring with the starting vector $\rket{s_0^z}$. 
The blue line is drawn at the numerically known dimension of the Krylov space $dim(\mathcal{K})=1893$.
}
\label{reortho-istf-perio}
\end{figure}

In Fig. \ref{reortho-istf-perio} it can be seen that the reorthogonalized Lanczos 
sequence does not terminate when the full dimension ($dim(\mathcal{K})=1893$, indicated by the blue vertical line) 
of the Krylov space is reached. 
To further investigate whether a vector $\rket{O_n}$ lays partially outside of the Krylov space one can consider the decomposition of $\rket{O_n}$ into the $\rket{\bar{E}_p^{\mathcal{L}}}$ 
\begin{align}
    \label{eq:deceig}
    \rket{O_n} = \sum_{p=1}^{dim(\mathcal{K})} d_p^n \rket{\bar{E}_p^{\mathcal{L}}}
    \ .
\end{align}
Since this also builds an ONB for the Krylov space, one can check 
whether the overlap of the newly generated vector $\rket{O_n}$ with the given ONB $\{ \rket{\bar{E}_p^{\mathcal{L}}} \}$
is still unity, 
thus showing that the operator is still completely contained within the Krylov space.
The overlap with the Krylov space is given by
\begin{align}
    C_n =\sqrt{\sum_{p=1}^{dim(\mathcal{K})} (d_p^n)^2}= 
    \sqrt{\sum_{p=1}^{dim(\mathcal{K})} \rbraket{O_n}{\bar{E}_p^{\mathcal{L}}}^2}
    \ .
\end{align}
This is shown by the green curve in Fig. \ref{reortho-istf-perio}.

Since a drop of $C_n$ at around $n=100$ can be observed it shows that the Lanczos sequence is no longer completely contained within the Krylov space quite early compared to the predicted dimension.
This shows that in general the Krylov space is numerically not closed under the Lanczos algorithm. 
It should be noted at this point that the diagonalization as well as the decomposition of the operators 
into the eigenbasis of $\mathcal{L}$ is done numerically.
Therefore, in section \ref{sec:dc} we will present a case where the eigenbasis is known analytically
and artifacts of a numerical diagonalization can be excluded.

\changed{In order to estimate the impact of the numerical accuracy on the
calculation presented in \figref{reortho-istf-perio} we repeated the calculation 
originally done in C++ with \emph{double} floats now with \emph{float} as well as 
\emph{long double}. The example showed that \emph{float} is absolutely untrustworthy; 
it produced much larger errors and a massively too large 
estimate of the dimension. This should be a warning to everyone still 
using graphics cards with float precision. The more precise calculations
with \emph{long double} yielded the same predicted dimension as for \emph{double}, 
similar $b_n$, and slightly later dropping $C_n$
(only 10 more steps). 
However, the usage of \emph{long double} allows for more precise 
tuning of the tolerance parameter in distinguishing degeneracies of the Liouvillian.}

\begin{figure}[ht!]
\centering
\includegraphics*[width=0.9\columnwidth]{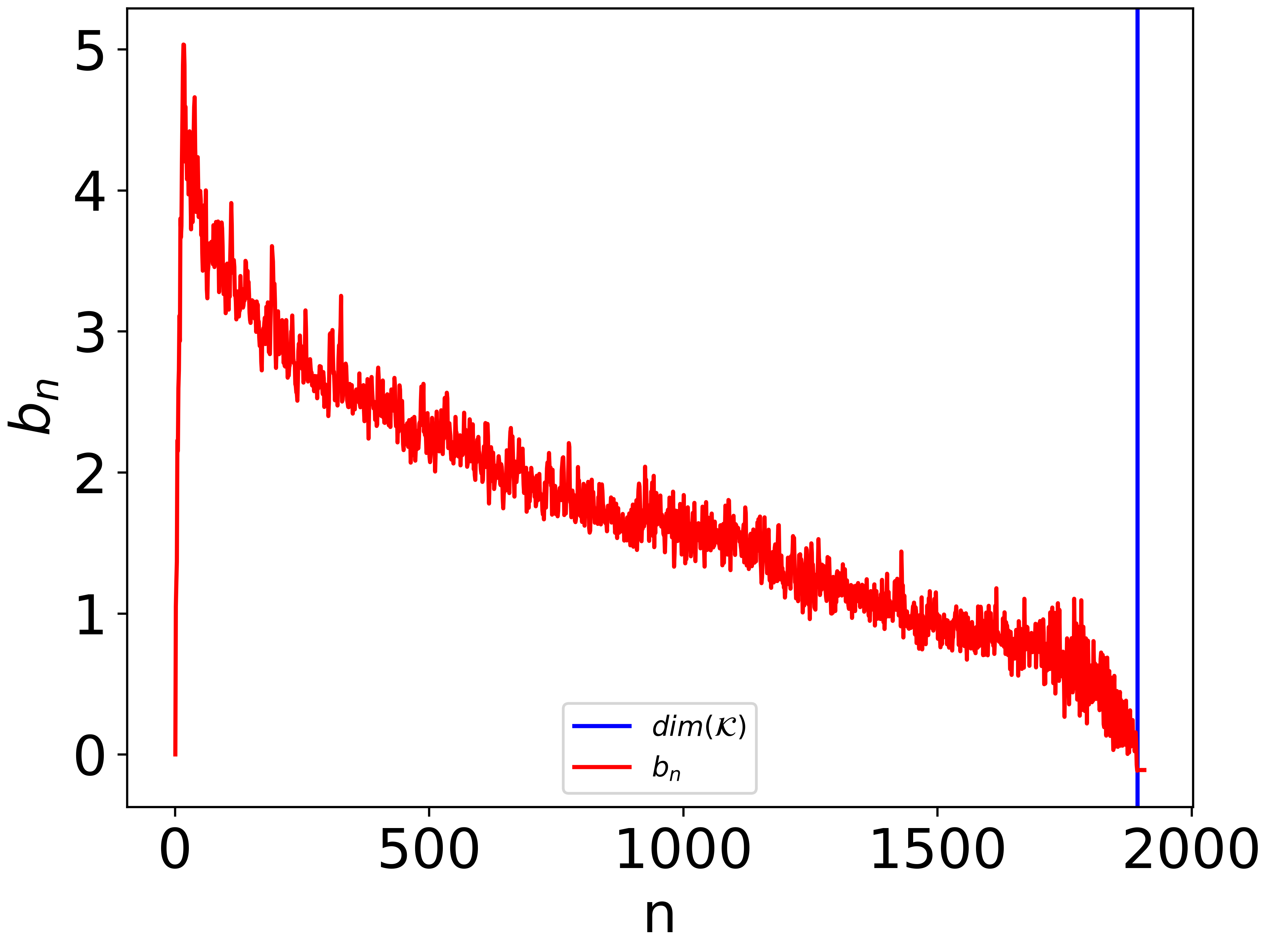}
\caption{Lanczos sequence $b_n$  with double reorthogonalization and a representation constrained to the Krylov subspace 
for the tilted-field Ising ring with the starting vector $\rket{s_0^z}$. 
The blue line is drawn at the numerically known 
dimension of the Krylov space $dim(\mathcal{K})=1893$.
}
\label{reortho-istf-perio-eig}
\end{figure}

Using a representation of the Liouvillian and the Lanczos vectors $\rket{O_n}$ 
which is constrained to the Krylov space can remove the possibility of leaving $\mathcal{K}$ during the Lanczos procedure.
This is done by considering the decomposition Eq. (\ref{eq:deceig}) and calculating the matrix representation of $\mathcal{L}$ via 
\begin{align}
\label{eq:constr-rep}
   \mathcal{L}_{k,k'}^{\mathcal{K}} = \rbra{\bar{E}_k^{\mathcal{L}}} \mathcal{L} \rket{\bar{E}_{k'}^{\mathcal{L}}}
   \ .    
\end{align}
Obviously, this is a diagonal matrix, however, it is not the entire diagonal matrix of $\mathcal{L}$ 
in the complete Hilbert space, since it does not contain the degeneracies of eigenvalues 
or eigenspaces that do not appear in the decomposition of the starting vector.
Doing the Lanczos iterations with this $dim(\mathcal{K}) \times dim(\mathcal{K})$ matrix 
and the $dim(\mathcal{K})$-dimensional vector following from Eq. (\ref{eq:deceig}) 
then leads to \figref{reortho-istf-perio-eig}.

\begin{figure}[ht!]
\centering
\includegraphics*[width=0.9\columnwidth]{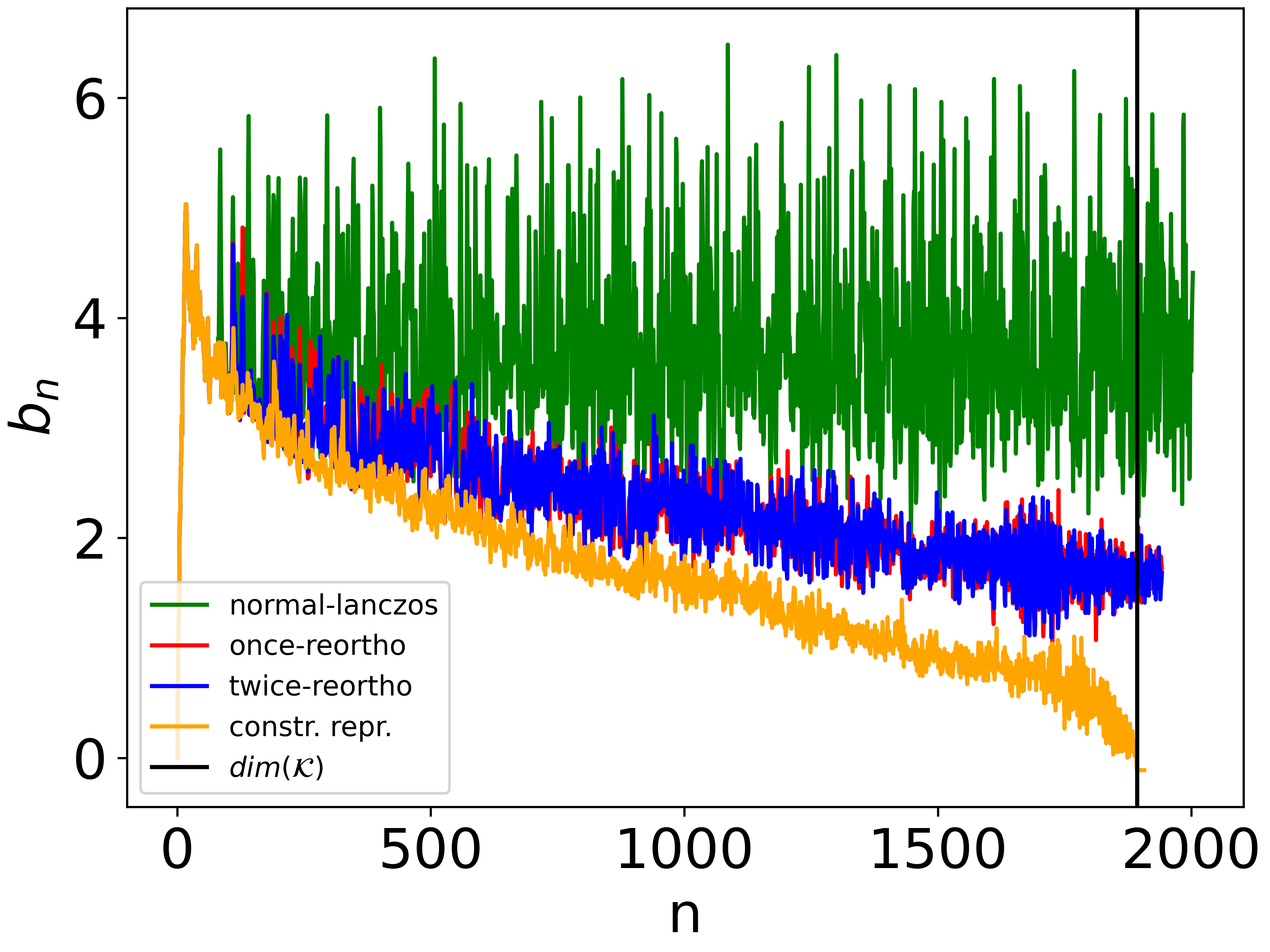}
\caption{Lanczos sequence $b_n$ calculated with different methods for the 
tilted field Ising ring with the starting operator $s_0^z$. 
The constrained representation (constr. repr.) means a representation of the Liouvillian according to Eq. (\ref{eq:constr-rep}).
The black 
line is drawn at the calculated dimension of the Krylov space $dim(\mathcal{K})=1893$.
}
\label{istf-lanc-vergl}
\end{figure}

Now the Lanczos sequence terminates at the Krylov space dimension. 
This is no surprise, since we chose a representation which enforces this behavior. 
Nonetheless, we eliminated the effect of leaving the Krylov space during the Lanczos procedure. 
It should be noted however that a lot of extra computational effort had to be made for this, 
because we need to diagonalize $\mathcal{L}$, 
represent our starting operator with respect to its eigenvectors $\rket{E_k^{\mathcal{L}}}$
and numerically distinguish non-degeneracy of the eigenvalues.
Even though this way the resulting $b_n$ seem better, 
in the sense that there are no artifacts of leaving the Krylov space, 
we still can not say that they are definitely closer to the 'true' $b_n$ 
at all iteration steps.
In Fig. \ref{istf-lanc-vergl} the Lanczos sequence is shown for different methods of calculation. 
One can see that for the single and double reorthogonalization 
the $b_n$ are closer to the case of a constrained representation. 
However, the problem of non-vanishing $b_n$ at $dim(\mathcal{K})$ is still visible.

\subsection{One magnon space of the Heisenberg delta chain \label{sec:dc}}

\begin{figure}[ht!]
\centering
\includegraphics*[width=0.9\columnwidth]{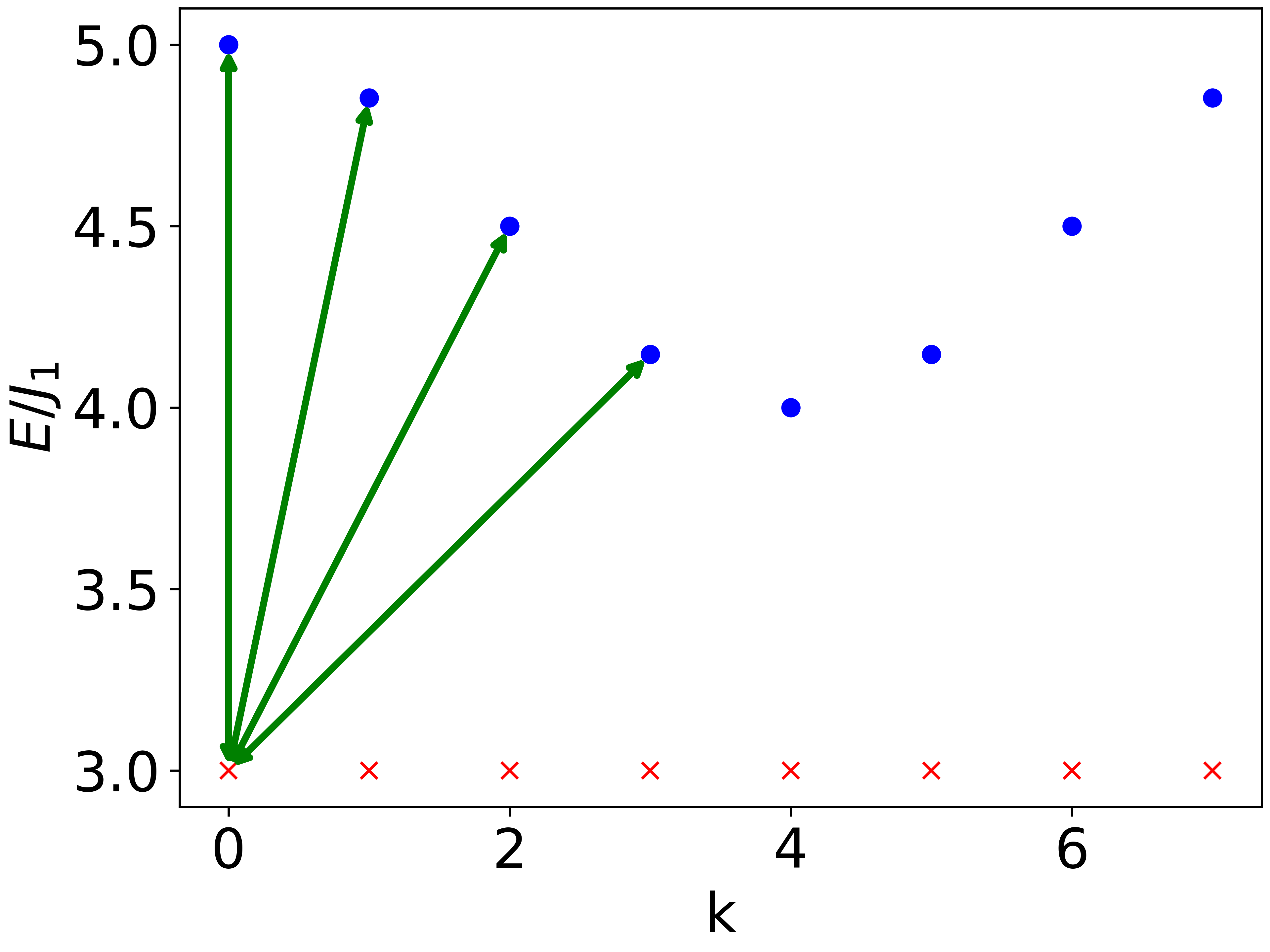}
\caption{One magnon band structure of the Heisenberg delta chain 
for $N=16$ and choice of eigenvectors used for the tensor product.
Note that the arrows point in both directions.
}
\label{tensprod-ana}
\end{figure}

All of this can already be seen when considering a small symmetry-related 
subspace of a Hilbert space.
For this purpose, we consider the ($M=Ns-1$)-subspace of the antiferromagnetic 
Heisenberg delta chain given by the Hamiltonian
\begin{align}
    H = J_1 \sum_{i=0}^{N-1} \vec{s}_i \cdot \vec{s}_{i+1} + 
J_2 \sum_{i=0}^{\frac{N}{2}-1} \vec{s}_{2i} \cdot \vec{s}_{2i+2}
\ ,
\end{align}
which has a flat band for the case of $J_2/J_1=1/2$, see \cite{JES:PRB23} and references therein.
The Hamiltonian in this small subspace can be diagonalized analytically leading to the eigenvalues of $H$
\begin{align}
    E^H_0&=\frac{NJ_1}{4}+\frac{NJ_2}{8}-4J_2 \\ 
    E^H_1&=\frac{NJ_1}{4}+\frac{NJ_2}{8} - J_2 \left[1-\cos \left(\frac{4 \pi k}{N} \right) \right],
\end{align}
which can be used to calculate the eigenvalues of the Liouvillian via (\ref{eq:ham-liou}).
The eigenvectors of $H$ and thus of $\mathcal{L}$ are also known analytically.
If one takes as a starting operator a chosen superposition of these eigenvectors, 
it is no longer a question of numerical accuracy to find the minimal number of 
eigenvectors of $\mathcal{L}$ needed to represent the starting operator.
To make sure we do not use degenerate eigenvectors we only use those constructed by the tensor product $\ket{E^H_{0,k=0}}\bra{E^H_{1,k'}} = \rket{E^{\mathcal{L}}_m}$ with $k'<\frac{N}{4}$ (see Fig. \ref{tensprod-ana}), 
thus leading to a strictly monotonic increase in $E^{\mathcal{L}}_{m(k-k')}=E^H_{0,k=0}-E^H_{1,k'}$ with $k'$.
In addition to these eigenvectors, we also include the adjoint of every vector $\ket{E^H_{1,k'}}\bra{E^H_{0,k=0}} =\rket{E^{\mathcal{L}\dagger}_m}$ in order to later construct a hermitian operator as a starting vector in this subspace. These adjoined eigenvectors are not degenerate, as adjoining the vectors introduces a sign change in their energies $E^{\mathcal{L}}_{m(k'-k)} = - E^{\mathcal{L}}_{m(k-k')}$.

\begin{figure}[ht!]
\centering
\includegraphics*[width=0.9\columnwidth]{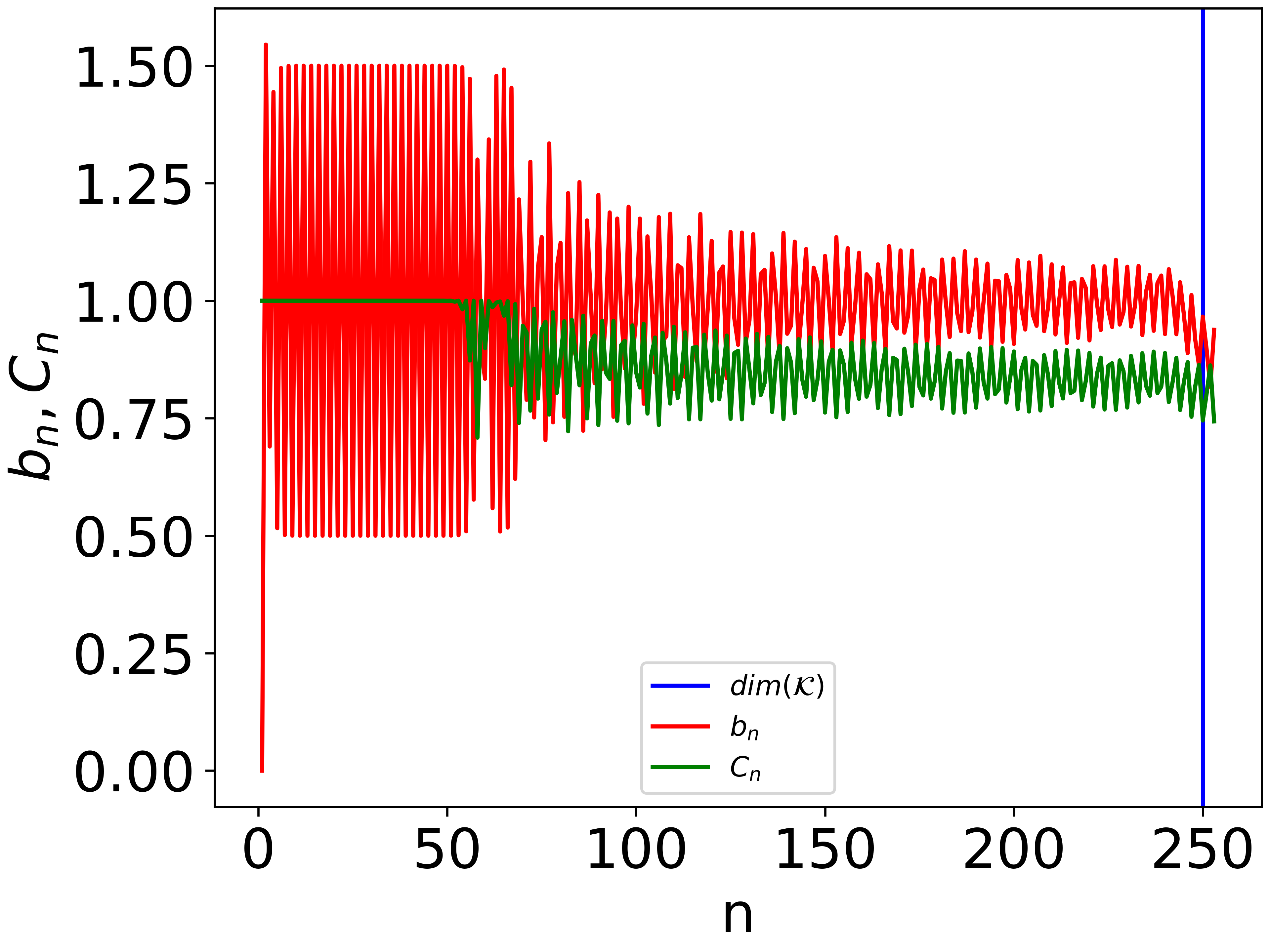}
\caption{Lanczos sequence $b_n$ and overlap $C_n$ for the
one-magnon space of the Heisenberg delta chain with the starting vector $\rket{A}$. 
The blue line is drawn at the analytically known dimension of the Krylov space $dim(\mathcal{K})=250$.
}
\label{reortho-lanczosana-500}
\end{figure}

Choosing now a uniform superposition of these eigenvectors and their adjoint counterparts
\begin{align}
    \rket{A}=\sqrt{\frac{4}{N}} \sum_{m=0}^{N/4-1} ( \rket{E^{\mathcal{L}}_m}  +\rket{E^{\mathcal{L}\dagger}_m} )
\end{align}
in an example case of $N=500$ leads to an analytically known dimension and basis 
of the Krylov space of $dim(\mathcal{K})=250$.
It should be noted that the operator we constructed this way 
is hermitian but not necessary local or of any obvious physical meaning.
However, one can still execute the Lanczos algorithm and investigate if the sequentially 
constructed vectors partially leave the Krylov space. 
Computing the Lanczos algorithm for this system and starting vector 
leads to the sequence shown in \figref{reortho-lanczosana-500}.

Again it is visible that the vectors leave the Krylov space 
at around $n=55$
where $C_n$ drops from one.
The $b_n$ show some strongly alternating behavior, 
\changed{It could be speculated that these oscillations are 
related to the flat band in our example. We refer interested readers to 
\cite{WLS:PRB24}.
But since we do not consider a local or even physical starting operator 
the alternating behavior is not necessarily of any bigger meaning.}

\section{Summary}

In this paper we showed that vectors generated numerically via the Lanczos algorithm 
do not need to reside in the Krylov space of the starting vector. 
This might even be of advantage for some applications like finding the eigenvalues of an operator, 
since as shown in appendix \ref{app:a} even when starting with a vector 
which does not have an overlap with the ground state 
it might still be reached numerically. 
However, in situations where one fine tunes this starting vector to only have an overlap with some of the eigenvectors
of the considered Hamiltonian or Liouvillian it is not guaranteed that it will stay inside the corresponding Krylov space. 
Especially for the operator Lanczos discussed above, where one chooses a physically meaningful operator 
as a staring vector and cannot chose the number of eigenspaces needed for its eigendecomposition 
this can play a role in the behavior of the Lanczos sequence, 
in particular for the point, where the Lanczos sequence should terminate. 
To investigate this we looked at the tilted-field Ising ring with a local operator as the starting vector.
For this example, the starting vector could only be decomposed into eigenvectors of the Liouvillian numerically. 
To eliminate artifacts due to numerical diagonalization we included an example 
of a small subspace of the Heisenberg delta chain where the eigendecomposition 
of the starting vector is known analytically. 
In both cases, the problem of leaving the Krylov space was observable. 

\changed{Our own experience suggests that the observed behavior also occurs for
schemes that employ Pauli strings \cite{Pieper:M25}. In these schemes, much larger systems
can be modelled, however, due to the exponential increase 
in required computational power and memory the number of achievable iterations 
is limited to the 15-40 iterations for instance achieved in 
\cite{PhysRevX.9.041017}. It may thus be that the discussed numerical problems
which should also occur for the Lanczos coefficients in schemes using Pauli strings
are not yet observable at the lower number of iterations.}

\changed{In addition, it should be noted that the discussed numerical instability
also occurs in the usual Lanczos algorithm as demonstrated in the appendix.}

\section*{Acknowledgment}

This work was supported by the Deutsche Forschungsgemeinschaft DFG
(355031190 (FOR~2692); 397300368 (SCHN~615/25-2)). 
We thank Coraline Letouzé, Jiaozi Wang and Jochen Gemmer for valuable discussions.

\appendix
\section{Example of leaving the subspace for ordinary state Lanczos \label{app:a}}

In this appendix, we like to show that the problem discussed above
arises in ordinary Lanczos procedures where an operator, e.g.\ the Hamiltonian 
is applied to a starting vector to generate a Krylov space of which
the smallest eigenvalue is taken as an approximation of the true 
smallest eigenvalue. The failure of the Lanczos procedure can easily 
be provoked by choosing a starting vector that belongs to an
invariant subspace due to a symmetry of the Hamiltonian. If the 
calculation is not restricted to this subspace but performed in a
larger space the dynamics in finite-precision arithmetics generates
a sequence of Lanczos vectors that escape the invariant subspace 
eventually.

\begin{figure}[ht!]
\centering
\includegraphics*[width=0.9\columnwidth]{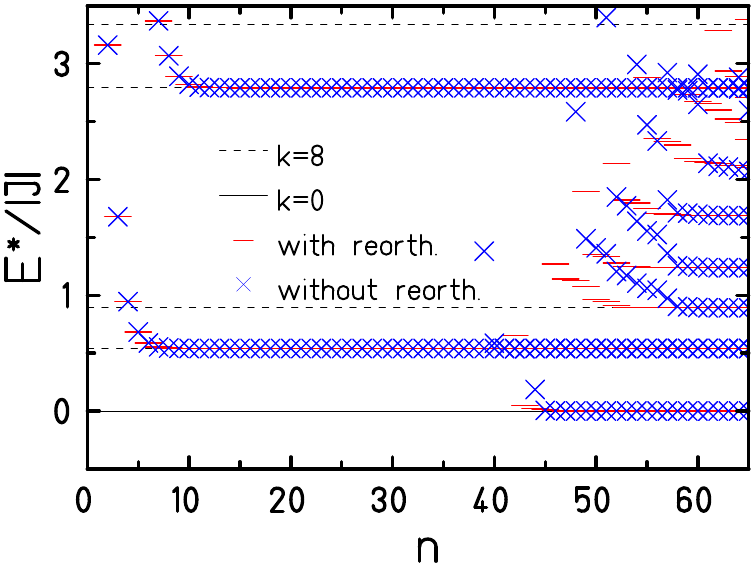}
\caption{Sequence of (low-lying) energy eigenvalues of $T$ 
at every Lanczos step $n$ in the vector space with $M=0$. 
The dimension of the subspace with $M=0$ and $k=8$
is 810, taking additionally into account the spin flip symmetry, the dimension 
of the subspace with additional odd spin-flip symmetry is 396. The starting 
vector has got odd symmetry, the ground state even.
The solid black line marks the exact ground state energy
for $k=0$; dashed lines mark exact eigenenergies for $k=8$.
The Lanczos sequence with reorthogonalization is depicted
by red dashes, the respective sequence without reorthogonalization 
by blue x-symbols.
}
\label{reortho-ring-16}
\end{figure}

In the specific example we consider a quantum spin ring with $N=16$
and $s=1/2$ in the Heisenberg model with the same exchange interaction $J$
between nearest neighbors ($-2J$-convention). This system has got several symmetries:
$SU(2)$ since it commutes with all components of the total spin, 
translational symmetry along the ring expressed as $C_N$ point group
as well as spin flip symmetry in the subspace with total magnetic 
quantum number $M=0$ \cite{BSS:JMMM00}.
We perform the calculation in the subspace with $M=0$
and choose as starting vector $1/\sqrt{2}(\ket{1010101010101010}-\ket{0101010101010101})$, 
which is a superposition of two product states where $0$ stands for a local $m_s=1/2$
and $1$ for $m_s=-1/2$. This vector is an eigenstate of the shift operator 
($T\equiv C_{16}$)
with shift quantum number $k=8$, whereas the true ground state possesses 
$k=0$ \cite{BSS:JMMM00:B}. An exact Lanczos procedure should generate a sequence of states 
that respects the good quantum number $k=8$. 

\begin{figure}[ht!]
\centering
\includegraphics*[width=0.9\columnwidth]{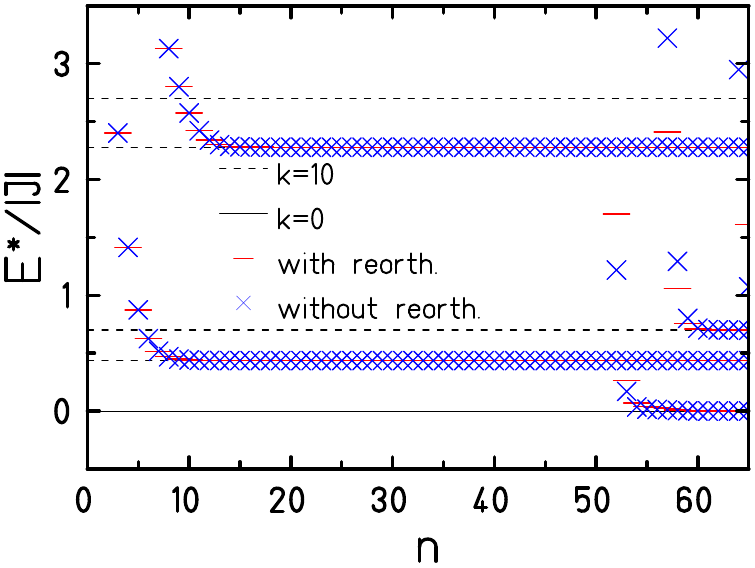}
\caption{Sequence of (low-lying) energy eigenvalues of $T$ 
at every Lanczos step $n$ in the vector space with $M=0$. 
The dimension of the subspace with $M=0$ and $k=10$
is 9252, taking additionally into account the spin flip symmetry, the dimension 
of the subspace with additional odd or spin-flip symmetry is about half.
The starting 
vector has got odd symmetry, the ground state even.
The solid black line marks the exact ground state energy
for $k=0$; dashed lines mark exact eigenenergies for $k=10$.
The Lanczos sequence with reorthogonalization is depicted
by red dashes, the respective sequence without reorthogonalization 
by blue x-symbols.
}
\label{reortho-ring-20}
\end{figure}

However, as \figref{reortho-ring-16} demonstrates, in a computer simulation 
the numerical sequence acquires contributions from states with e.g.\ $k=0$. 
These contributions grow exponentially fast with Krylov space dimension,
in particular, in cases they contain the ground state. In the example
simulation this happens after about 40 Lanczos steps. Before, the algorithm 
has accurately found the lowest energy compatible with $k=8$, but then it
develops a component along the overall ground state that belongs to $k=0$,
and within a few steps converges also to the respective eigenvalue.
This happens with (red symbols) and and without (blue symbols) 
reorthogonalization, which once more strengthens our statement that 
reorthogonalization does not guard against components orthogonal to the
true/exact Krylov space. 

With \figref{reortho-ring-20} we want to demonstrate that the problem
arises after approximately the same number of Lanczos steps and in particular
does not scale with the dimension of the underlying spaces. For the example of 
$N=20$ spins $s=1/2$ the dimensions are about ten times larger, but the 
escape of the Lanczos sequence into the orthogonal complement of the true
Krylov space happens after about 50 steps compared to 40 for $N=16$, 
see \figref{reortho-ring-16}.

\begin{figure}[ht!]
\centering
\includegraphics*[width=0.9\columnwidth]{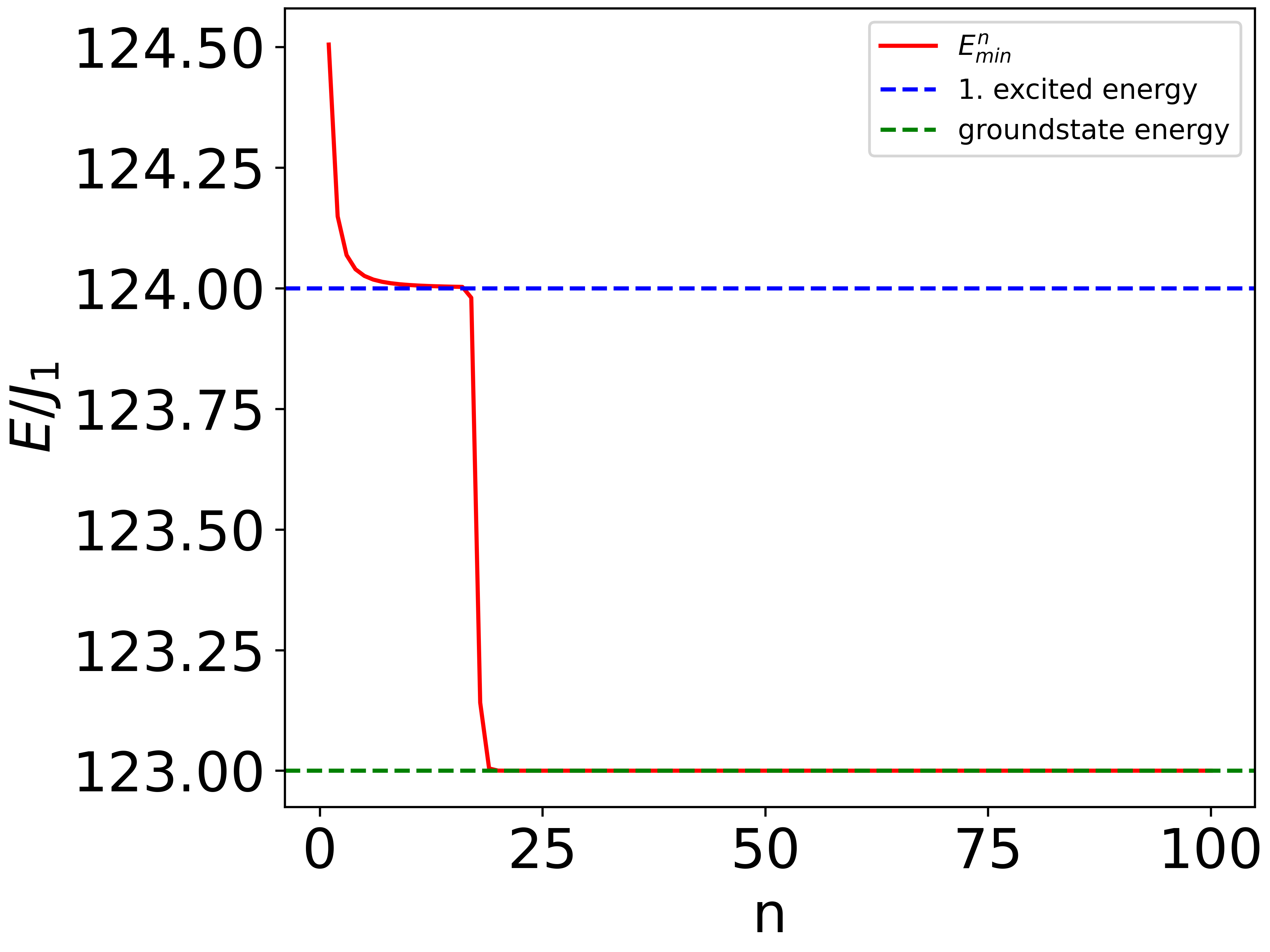}
\caption{Minimal energies of the tridiagonal matrix for each Lanczos step 
in the one-magnon subspace of the Heisenberg delta chain with $N=400$. 
The analytically known dimension of the Krylov space is $dim(\mathcal{K})=101$. 
The blue and green dashed lines indicate the energy of the first excited 
and ground state energy respectively.
}
\label{statelanconemag}
\end{figure}

To further strengthen this finding we again considered the one-magnon subspace of 
the Heisenberg delta chain (compare section \ref{sec:dc}). 
Since we investigate the ordinary state Lanczos we just superimposed 
all states from the upper band with $k\leq N/4$ for our starting vector,
compare \figref{tensprod-ana}.
Running the Lanczos algorithm and diagonalizing the tridiagonal matrix 
in every step leads to the minimal energies seen in
\figref{statelanconemag}.
After the minimal energy seems to saturate at the first excited state, as it should, 
at around $n=19$ it quickly drops to the ground state energy, 
even though the initial state should has no overlap with the ground state.

In typical applications in quantum magnetism this problem does not occur because
one restricts the vector space to the smallest possible space for the given 
symmetries. Then, a Lanczos procedure cannot leave this vector space by construction.


%

\end{document}